\documentclass[11pt]{article}
\usepackage[utf8]{inputenc}
\usepackage[a4paper, total={6in, 8in}]{geometry}
\usepackage{graphicx}

\usepackage{amsmath}
\usepackage{comment}
\usepackage{amsthm}

\usepackage{xcolor}
\usepackage{hyperref}
\hypersetup{
    colorlinks=true,
    linkcolor=brown,
    filecolor=brown,  
    citecolor=brown
}

\newtheorem{theorem}{Theorem}
\newtheorem{lemma}{Lemma}

\newtheorem{proposition}{Proposition}

\theoremstyle{definition}

\renewcommand{\title}[1]{\vbox{\center\LARGE{#1}}\vspace{5mm}}
\renewcommand{\author}[1]{\vbox{\center#1}\vspace{5mm}}

\newcommand{\address}[1]{\vbox{\center\em#1}}

\DeclareMathOperator{\tr}{\text{tr}}

\begin{document}

\title{Quantum Cross Entropy and Maximum Likelihood Principle}

\begin{center}
\vspace{5mm}
Zhou Shangnan,$^{a}$\footnote{\tt snzhou@stanford.edu.} and Yixu Wang$^{b}$ 
\end{center}

\address{${}^a$Stanford Institute for Theoretical Physics, \\
Stanford University, Stanford, CA 94305, USA}
\address{${}^{b}$Maryland Center for Fundamental Physics, \\
and Department of Physics, University of Maryland, College Park, MD 20742, USA}

\date{\today}
\vspace{5mm}

\begin{abstract}
    Quantum machine learning is an emerging field at the intersection of machine learning and quantum computing. Classical cross entropy plays a central role in machine learning. We define its quantum generalization, the quantum cross entropy, prove its lower bounds, and investigate its relation to quantum fidelity. 
    In the classical case, minimizing cross entropy is equivalent to maximizing likelihood. 
    In the quantum case, when the quantum cross entropy is constructed from quantum data undisturbed by quantum measurements, this relation holds. 
    Classical cross entropy is equal to negative log-likelihood. 
    When we obtain quantum cross entropy through empirical density matrix based on measurement outcomes, the quantum cross entropy is lower-bounded by negative log-likelihood. 
    These two different scenarios illustrate the information loss when making quantum measurements. We conclude that to achieve the goal of full quantum machine learning, it is crucial to utilize the deferred measurement principle.
\end{abstract}

\section{Introduction}

Machine learning has attracted interests from various fields as a powerful tool for finding patterns in data. Supported by machine learning technology, computer programs can improve automatically through experience, which has enabled a wide spectrum of applications: from visual and speech recognition, effective web search, to study of human genomics \cite{lecun2015deep,alphafold}. Classical machine learning techniques have also found many interesting applications in different disciplines of quantum physics \cite{MLtomo,manybody,phasematter,QMBstateDNN,MLPhaseTrans,MLTopoState,MLNNQST,MLdeepmatrices}.

With the advancement of quantum information science and technology, there is both theoretical and practical interest in understanding quantum systems, building quantum devices, developing quantum algorithms, and ultimately, taking advantage of quantum supremacy \cite{supremacy1,supremacy2}. Quantum machine learning overlaps with the current interests in quantum computing and machine learning in many different aspects. 
For supervised and unsupervised machine learning tasks, we are usually given a set of training data and a set of test data. We process the training data using our learning algorithm, so that we can form a prediction model. Then we use the prediction model to process the test data and give predictions. The learning algorithm usually consists of several steps. Hence, there can be different levels of quantization. At the lowest level, we can use classical processors and classical learning algorithms to study quantum systems \cite{MLtomo,manybody,phasematter,QMBstateDNN,MLPhaseTrans,MLTopoState,MLNNQST,MLdeepmatrices}. Moreover, we can transform part or all of the learning algorithm to be quantum, so that we can speed up the learning process or improve the learning performance \cite{lloyd2013quantum, quantumpca,huang2021information, KernelNature, KernelPRL}. The ultimate goal is to achieve full quantum machine learning, where the input data and output predictions, the processing device, and the learning algorithms, are all fully quantum and enjoy quantum supremacy. 
While there have been many attempts in designing quantum machine learning architectures \cite{deepqnn,cao2017quantum,wecker2016training,killoran2019continuous,arrazola2019machine,steinbrecher2019quantum}, the theoretical aspects of quantum machine learning are still lacking.

In this work, we study the quantum generalizations of the cross entropy and maximum likelihood principle. Cross entropy is widely adopted as a loss function for machine learning tasks. To achieve full quantum machine learning, we have to process quantum data and quantum training procedures, which makes a quantum version of cross entropy desirable. We believe that with a loss function more closely related to the quantum nature of quantum systems, there is a large sea of possible quantum machine learning algorithms that can provide speed-ups or enhance performance for various tasks. At the same time, previous works \cite{jozsa1994fidelity,petz2001entropy,relative1,relative2,casini2008relative,gu2010fidelity,muller2013renyi,jafferis2016relative, berta2017variational, shangnan2019complexity} show that the study of a quantum version of an important classical quantity itself is interesting and opens doors for many applications in the quantum regime.
Many supervised, unsupervised, and semi-supervised machine learning algorithms, including but not limited to logistic regression, Gaussian discriminant analysis, and the Expectation-Maximization algorithm, are based on the maximum likelihood principle. The principle tells us to estimate the parameters of a statistical model by maximizing a likelihood function, so that the observed data is most probable. 
Classically, minimizing the cross entropy is the same as maximizing the likelihood. More concretely, classical cross entropy is equal to average negative log-likelihood. It is natural to ask: do these relations hold in the quantum case? If they don't always hold, what can we learn from that? Our work answers these questions.

In Section \ref{qce}, we define a new quantum information quantity, the quantum cross entropy $S(\rho, \sigma) =  -\tr(\rho \log \sigma)$, as a direct quantum generalization of the classical cross entropy. We proved its positivity, invariance under unitaries, linearity on the first argument, convexity on the second argument, and joint convexity.
We also show that it is lower-bounded by the negative log fidelity, namely $S(\rho, \sigma) \geq -\log \tr (\rho \sigma)\geq -\log F(\rho,\sigma)$.
In Section \ref{cce}, we review the classical cross entropy, the maximum likelihood principle, and their interplay.
In Section \ref{qcelllh}, we examine quantum cross entropies constructed from different formats of quantum data. When the quantum data used for quantum cross entropy is undisturbed by quantum measurements, minimizing the quantum cross entropy is the same as maximizing the likelihood function. Unfortunately, this doesn't hold when measurements come into play. However, we still have something worth-noting when our quantum cross entropy is reconstructed from the empirical density matrix, a quantum generalization of the empirical probability distribution. In this case, quantum cross entropy is lower-bounded by negative log-likelihood. The inequality, as well as the differences between two scenarios, shows that quantum measurements, and the information loss comes with it, breaks some nice properties that hold in the classical case. With this, we propose a guiding principle for doing full quantum machine learning, which is that we should maximize the usage of deferred measurement principle when designing quantum algorithms, to avoid strange behaviors \cite{Barren} in quantum machine learning.

\section{Quantum cross entropy}\label{qce}

Despite the fact that classical cross entropy, classical and quantum relative entropies are defined and studied, a formal definition of quantum cross entropy is lacking. In this section, we first review the definition of classical cross entropy, and show its connection to classical fidelity. We then generalize this concept to the quantum case, and present properties and physical intuitions of the quantum cross entropy.

In classical information theory, the cross entropy is the average number of bits needed to encode data coming from a source with distribution $p$ when we use model $q$ to define our code-book. 
Formally, the classical cross entropy of the distribution $q$ relative to a distribution $p$ over a given set is defined as \cite{MLbookcross}

\begin{equation}
    H(p,q) = -E_p [\log q],
\end{equation}
where $E_p$ is the expectation value operator with respect to the distribution $p$. 

For discrete probability distributions $p$ and $q$ with the same support $\mathcal{X}$, the classical cross entropy is

\begin{equation}
    H(p,q) = -\sum_{x \in \mathcal{X}} p(x) \log q(x).
\end{equation}

Equivalently, it can be written as the sum of Shannon entropy $H(p)$ and KL divergence from $p$ to $q$ (also known as classical relative entropy of $q$ with respect to $p$):

\begin{equation}
\label{classical entropy}
    H(p,q) = H(p) + D_{KL} (p \| q).
\end{equation}

%Here, we focus on the case when we have discrete probability distributions, partly because later we study finite-dimensional density matrices as a quantum generalization. It is expected that similar results follow for the continuous case.

Now we show that classical cross entropy has lower bounds:

\begin{equation}
    H(p,q) \geq -\log \Big(\sum_{x \in \mathcal{X}}
    p(x) q(x) \Big) \geq - \log F(p,q),
\end{equation}
where $F(p,q) = \Big(\sum_{x \in \mathcal{X}} \sqrt{p(x) q(x)}\Big)^2$ is the classical fidelity.

\begin{proof}
The first inequality is a consequence of concavity of the log function. For the second inequality, let $h(x) = p(x) q(x)$, it is enough to show that 
\begin{equation}
    \sum_{x \in \mathcal{X}} h(x) \leq \Big(\sum_{x \in \mathcal{X}} \sqrt{h(x)} \Big)^2,
\end{equation}
which is manifest.
\end{proof}

We define the quantum cross entropy by extending the classical definition from probability distributions to density matrices. For two density matrices $\rho$ and $\sigma$, the quantum cross entropy is

\begin{equation}
    S(\rho, \sigma) = -\tr(\rho \log \sigma),
\end{equation}
if the support of $\rho$ is contained in the support of $\sigma$, and $+\infty$ otherwise.

Equivalently, it can be written as 
\begin{equation}
\label{quantum entropy}
    S(\rho, \sigma) = S( \rho \| \sigma) 
    - \tr(\rho \log \rho) = S(\rho \| \sigma) + S(\rho),
\end{equation}
where $S(\rho\|\sigma)$ is the quantum relative entropy, and $S(\rho)$ is the von Neumann entropy of $\rho$. This shows that our definition of quantum cross entropy is consistent with existing definitions, and Eq (\ref{quantum entropy}) has exactly the same form as Eq (\ref{classical entropy}). Hence, the quantum cross entropy is also a quantum distance measure. 
We conjecture that the quantum cross entropy is the average number of qubits needed to encode data from a quantum source represented by a density matrix $\rho$, when we use a rather universal compression protocol for the quantum source $\sigma$.

Now we summarize some properties of quantum cross entropy. 

\begin{proposition}
 $S(\rho, \sigma)$ is zero if and only if $\rho = \sigma$ and $\rho$ is a pure state, otherwise it is positive.
\end{proposition}

\begin{proposition}
 $S(\rho, \sigma)$ is invariant under simultaneous unitary transformation on both $\rho$ and $\sigma$:
\end{proposition}

\begin{equation}
     S(\rho, \sigma) = S(U \rho U^\dagger, U \sigma U^\dagger).
\end{equation}

\begin{proposition}[Linearity on the first argument]\label{linearity}

\begin{equation}
    S \Big(\sum_i p_i \rho_i, \sigma \Big) = 
    \sum_i p_i S(\rho_i, \sigma).
\end{equation}

\end{proposition}

\begin{proposition}[Convexity on the second argument]\label{convexity}

\begin{equation}
    S \Big(\rho, \sum_j q_j \sigma_j \Big)
    \leq \sum_j q_j S (\rho, \sigma_j ).
\end{equation}
\end{proposition}

\begin{proof}
This follows from the joint convexity of the relative entropy $S(\rho \| \sigma)$. When $\rho$ is a given density matrix, $-\tr(\rho \log \rho)$ is cancelled, so the remainder term $-\tr(\rho \log \sigma)$ is convex with respect to $\sigma$.
\end{proof} 

\begin{proposition}[Joint convexity]
\begin{equation}
    S \Big(\sum_i p_i \rho_i, \sum_j q_j \sigma_j \Big)
    \leq \sum_i \sum_j p_i q_j S(\rho_i, \sigma_j).
\end{equation}
\end{proposition}

This is a corollary of Properties \ref{linearity} and \ref{convexity}.

\begin{proposition}[Extensivity]\label{extensivity}

Let $\rho=\rho_1\otimes\rho_2...\otimes\rho_n$ and $\sigma=\sigma_1\otimes \sigma_2...\otimes\sigma_n$, then
\begin{equation}
S(\rho,\sigma)=\sum_{i=1}^nS(\rho_i,\sigma_i).
\end{equation}
    
\end{proposition}

\begin{proposition}\label{crossentropybound}
 $S(\rho, \sigma)$ has lower bounds:

\begin{equation}\label{cross entropy-fidelity}
    S(\rho, \sigma) \geq -\log \tr (\rho \sigma)\geq -\log F(\rho,\sigma),
\end{equation}
where $F(\rho,\sigma)= \big(\tr \sqrt{\rho^{1/2}\sigma\rho^{1/2}} \big)^2$ is the quantum fidelity between two states.
\end{proposition}

\begin{proof}
We work in finite dimensional Hilbert spaces so the density matrices are $n\times n$ Hermitian matrices. The condition $\text{supp}(\rho) \subseteq \text{supp}(\sigma)$ is necessary to have  $S(\rho,\sigma)$ well defined. If $\text{rank}(\sigma)=s\leq n$, we may reduce the problem to an $s$ dimensional Hilbert space. So without loss of generality, we take the projection $P_\sigma$ to $\text{supp}~\sigma$  as $P_\sigma=I_n$ in $n$-dimensional Hilbert space. 

If $\text{supp}(\rho) = \text{supp}(\sigma)$, then $F=\log\sigma$ and $R=\log \rho$ are well-defined $n \times n$ Hermitian matrices, so by Peierls-Bogoliubov inequality, together with the Golden-Thompson inequality,
\begin{equation}
    \tr(e^F e^R)\geq \tr(e^{F+R})\geq e^{\tr(Fe^R)}.
\end{equation}

We get the desired inequality by taking the logarithm of both sides.

If $\text{supp}(\rho) \subset \text{supp}(\sigma)$, the inequality can be obtained via a limiting procedure as follows.
Consider a sequence of density matrices ${\rho_n}$ such that $\text{supp}(\rho_n) = \text{supp}(\sigma)$ and
\begin{equation}
    \lim_{n\to\infty}\|\rho_n-\rho\|=0.
\end{equation}
Here $\|\cdot\|$ denotes the operator norm.
It is straightforward to show that 
\begin{equation}
    \lim_{n\to\infty}-\tr(\rho_n\log\sigma)=-\tr(\rho\log\sigma),\quad\lim_{n\to\infty}-\log\tr(\rho_n\sigma)=-\log\tr(\rho\sigma).
\end{equation} 
For the first one,
\begin{equation}
    0 \leq \lim_{n\to\infty}|\tr((\rho_n-\rho)\log\sigma)|\leq \|\log\sigma\|\lim_{n\to\infty}\|\rho_n-\rho\|=0.
\end{equation}
The other one can be proved similarly. As $\text{supp}(\rho_n) = \text{supp}(\sigma)$, we can apply the Peierls-Bogoliubov inequality as before to obtain
\begin{equation}
    -\tr(\rho_n\log\sigma)\geq-\log\tr(\rho_n\sigma).
\end{equation}
The existence of the limit under $n\to\infty$ on both sides implies the desired inequality when $\text{supp}(\rho) \subset \text{supp}(\sigma)$.

Now we investigate when the first inequality achieves equality $S(\rho,\sigma)=-\log\tr(\rho\sigma)$. The Golden-Thompson inequality holds if and only if when $F$ and $R$ commute. That is, when the density matrices $\rho$ and $\sigma$ can be simultaneously diagonalized. Intuitively, there exists a set of bases, in which both $\rho$ and $\sigma$ "look like classical probability distributions". Hence, it suffices to investigate the Peierls-Bogoliubov inequality when $\rho$ and $\sigma$ are classical distributions. This inequality, which is a corollary of the Klein's inequality, makes use of the convexity of the logarithmic function. The equality holds in either of the two cases. First, when $\sigma=e^g I$, that is, it is in a maximally mixed state, in other words, a uniform distribution. The $e^g$ factor can be fixed by the normalization condition of the state $\sigma$. Second, when $\rho$ commutes with $\sigma$ and $\text{rank}(\rho) = 1$.

The second inequality is a consequence of monotonicity of Schatten norm $\|T\|_1\geq \|T\|_2$, with positive semi-definite matrix $T = \sqrt{\rho^{1/2}\sigma\rho^{1/2}}$. The equality holds when $T$ is of rank 1. This is the case when $\rho$ and $\sigma$ are identical pure states.

\end{proof}

As an interesting side note, we give a simple scenario \cite{NielsenChuang} where the quantum cross entropy arises.
Suppose $\{Q_i\}$ is a complete set of projective measurement operators. If we measure state $\rho$, and do not read out the measurement result, then we obtain a state $\sigma = \sum_i Q_i \rho Q_i$. In this case, the quantum cross entropy is actually the von Neuman entropy of the after-measurement state $S(\rho, \sigma) = S(\sigma)$.

\section{Classical cross entropy and maximum likelihood principle}\label{cce}

One property of classical cross entropy that plays an important role in statistical learning theory, is that classical cross entropy is equal to average negative log-likelihood, also called the log loss. Hence, maximizing likelihood is equivalent to minimizing cross entropy. Many machine learning algorithms based on maximum likelihood principle use cross entropy as a cost function, especially when optimizing classification models. In this section, we review the argument and provide some additional insights from a measurement perspective.

First we introduce the concept of likelihood. We understand and predict the nature by constructing models, based on our observed data. For example, to estimate how biased a coin is, we basically want to learn the probability $\theta$ of getting a head when flipping it, and we call this $\theta$ our parameter.  We collect our data $\mathcal{D}$ by doing a sequence of identically and independently distributed coin flips and record the outputs:

\begin{equation}
    \mathcal{D} = x_1, x_2, ... , x_N,
\end{equation}
where each $x_i$ is either Head or Tail.

The probability of getting a certain $\mathcal{D}$, called the likelihood, is a function of parameter $\theta$. Formally, the likelihood function is 

\begin{equation}
    L(\theta) = P(\mathcal{D}|\theta) = \prod_{i = 1}^N
    P(x_i|\theta).
\end{equation}

The principle of maximum likelihood says that we should choose $\theta$ so as to make the data as highly probable as possible, which is to choose the $\theta$ that maximizes the likelihood function $L(\theta)$. The parameter $\theta$ also defines a probability distribution $P_{\theta}(x) = P(x|\theta)$, which tells us the probability of generating an output $x$ on a single trial.

Maximizing $L(\theta)$ is equivalent to maximizing any strictly increasing function of $L(\theta)$. In particular, the average log-likelihood $l(\theta)$ can make derivations much easier by turning the products into sums:

\begin{equation}
    l(\theta) = \frac{1}{N} \log P(\mathcal{D}|\theta) = \frac{1}{N} \sum_{i = 1}^N  \log P(x_i|\theta).
\end{equation}

The maximum likelihood estimator is the parameter which maximizes the (log) likelihood function:

\begin{equation}
    \theta_{MLE} = \arg \max_{\theta} l(\theta).
\end{equation}

In more general cases, $\theta$ can be a vector or matrix, or we can say we have multiple parameters $\theta_1$, $\theta_2$, ... , $\theta_m$. Our observed data can also be more complicated than a sequence of heads and tails, it can be a sequence of vectors or matrices. While the data and parameters become high-dimensional, the spirit of constructing a likelihood function and maximizing it remains the same.

From a physicist's perspective, we do a sequence of classical measurements to get some classical outputs, and we maximize the likelihood function to find the most probable parameters for our model. 

Now we show that classical cross entropy is equal to average negative log-likelihood. Note that the data points $x_i$'s and the parameter $\theta$ can both be seen as vectors. 

Suppose the probability distribution based on model parameter $\theta$ is $P_{\theta} (x) = P(x|\theta)$, where the support of $x$ is $\mathcal{X}$, and our data is $\mathcal{D} = \{x_i | 1 \leq i \leq N \}$. We can construct the "empirical data distribution":

\begin{equation}
    P_D(x) = \frac{1}{N} \sum_{i = 1}^N
    1\{x = x_i\},
\end{equation}
where the indicator function $1\{x = x_i\} = 1$ if $x = x_i$, and $1\{x = x_i\} = 0$ otherwise. This is a valid distribution since $\sum_{x \in \mathcal{X}} P_D(x)  = 1$.

The classical cross entropy $H\big(P_D(x), P_{\theta}(x)\big)$ is 

\begin{equation}
\begin{split}
    H\big(P_D(x), P_{\theta}(x) \big) 
    &=  -\sum_{x \in \mathcal{X}} P_D(x) \log P_{\theta} (x) 
    =  -\sum_{x \in \mathcal{X}} \frac{1}{N} 1\{x = x_i\} \log P (x | \theta) \\
    &=  -\sum_{i = 1}^N \frac{1}{N} \log P(x_i|\theta) 
    =  -l(\theta),
\end{split}
\end{equation}
which is equal to the average negative log-likelihood function $l(\theta)$.

Hence, minimizing classical cross entropy is equivalent to maximizing likelihood:

\begin{equation}
    \arg \min_{\theta} H\big(P_D(x), P_{\theta}(x)\big) =
    \arg \max_{\theta} l(\theta). 
\end{equation}

In practice, many machine learning algorithms based on maximum likelihood principle uses classical cross entropy as a cost function.

\section{The quantum generalization: the cost of quantum measurements}\label{qcelllh}

One necessary component of full quantum machine learning is to learn a quantum system from scratch. To do so, we need to gain further understanding of what "quantum data" means and how we obtain and manipulate them. To extract information from a quantum system, it is necessary to make measurements. However, we all know too well that when we make measurements, we disturb the quantum system and we lose some information of the original quantum system. Hence, it is crucial to examine the effects of measurements. 

The classical relation between the cross entropy and likelihood can be decomposed into two parts:

1. Minimizing cross entropy is maximizing likelihood.

2. Cross entropy is equal to negative log-likelihood.

Note that likelihood is based on classical probabilities, so we expect 2 to break down in the quantum case. Our conclusion is that, when the quantum cross entropy is from raw quantum data that is not affected by quantum measurements, 1 continues to hold. If we obtain the quantum cross entropy through empirical density matrix, which is constructed from measurement outcomes, then we have a modified version of 2: the quantum cross entropy is lower bounded by negative log-likelihood. The fundamental difference is the quantum data we use, which is a reflection of different levels of quantum measurements.

\subsection{Minimizing quantum cross entropy is maximizing likelihood}

Many quantum machine learning tasks involve learning a rather complex quantum state, fully or partially. Suppose we have a quantum state represented by density matrix $\rho$, with which we want to perform regression or classification tasks on, then naturally we put our model parameters into a density matrix $\sigma$, with restrictions based on prior knowledge of $\rho$.

In the ideal scenario, in which our quantum computer can calculate the quantum cross entropy directly without disturbing the states, then we have the perfect quantum data: quantum state $\rho$ that has not been measured or approximated. In this scenario, the quantum cross entropy used in calculation is $S(\rho, \sigma) = -\tr (\rho \log \sigma)$, and the model parameter that minimizes the quantum cross entropy is $\rho$:

\begin{equation}\label{minrho}
   \arg \min_{\sigma} S(\rho, \sigma) =  \arg \min_{\sigma} \big(-\tr (\rho \log \sigma) \big)
= \arg \min_{\sigma} S(\rho || \sigma) = \rho.
\end{equation}

To obtain the likelihood function, we make tomographic complete measurements $\{E_i\}$ on $\rho$. 
We may write $\{E_i\}$ as $\{E_{j k}\}$, where $j$ labels the $j$-th set of measurement operators, and $k$ labels the $k$-th measurement operator in the $j$-th set.  We have $\sum_k E_{j k} = I, \forall j$. We use the $j$-th set of measurement operators $N_j$ times, and the total number of measurements made is $N = \sum_j N_j$. We can write the average log-likelihood function $l(\sigma)$ by grouping the results into different sets:

\begin{equation}
    l(\sigma)=\frac1N\sum_{i=1}^N\log\tr(E_i\sigma)=\frac1N\sum_{j}\sum_{k}N_{j k}\log\tr(E_{j k}\sigma).
\end{equation}

The probability of having the $k$-th outcome using the $j$-th set of measurements is $p_{j k}\equiv \tr (E_{j k} \rho)$. In the limit $N_j \to\infty$ for all $j$, by the law of large numbers, we have

\begin{equation}
    \lim_{N_j\to\infty}\frac{N_{j k}}{N}=\frac{N_{j}}{N}\tr(E_{j k}\rho).
\end{equation}

Defining $q_{j k}\equiv \tr(E_{j k}\sigma)$, 
$\frac{N_j}{N}\equiv n_j$, the log-likelihood now takes the form
\begin{equation}
    l(\sigma)=\sum_{j}n_j\sum_{k}p_{j k}\log q_{j k}.
\end{equation}

Now we perform the variation to $l(\sigma)$ together with the sets of constraints that $\forall j,~\sum_{k}q_{j k}=1$. The extremal condition gives
\begin{equation}
    \delta\left(\sum_{j}n_j\sum_{k}(p_{j k}\log q_{j k}-\lambda_j q_{j k})\right)=\sum_{j}n_j\sum_{k}\left(\frac{p_{j k}}{q_{j k}}-\lambda_j\right)\delta q_{j k}=0. 
\end{equation}
Here $\delta q_{j k}\equiv\tr(E_{j k}\delta\sigma)$. $n_j$'s are completely arbitrary because we are free to choose the proportion of each set of measurements. The variational condition gives $\lambda_j q_{j k}=p_{j k}$. However, the normality condition $\sum_k p_{j k}=1$ fixes $\lambda_j=1$ and gives $q_{j k}=p_{j k}$. As $\{E_{j k}\}$ represents a tomographic complete set of measurements, we conclude that $\rho = \sigma$, and $\rho$ can be calculated through linear inversion. Hence, that model parameter that maximizes the log-likelihood is $\rho$:

\begin{equation}\label{maxrho}
    \arg \max_{\sigma} l(\sigma) = \rho.
\end{equation}

Combining Eqs (\ref{minrho}) and (\ref{maxrho}), we conclude that in this scenario, minimizing quantum cross entropy is maximizing likelihood:

\begin{equation}
    \arg \min_{\sigma} S(\rho, \sigma) = \arg \max_{\sigma} l(\sigma).
\end{equation}

\subsection{Quantum cross entropy is lower-bounded by negative log-likelihood}

In real life, fine-tuning a quantum circuit so that we can vary $\sigma$ and calculate $S(\rho, \sigma)$ is hard, if not impossible. For practical purposes, we look at another approach, where we generate rather "dirty" quantum data through measurements. To do so, we borrow the ideas of constructing empirical data distribution from classical statistics.

In the classical case, we generate a bunch of data by doing a bunch of independent classical measurements. In the quantum case, our data generating process will be doing quantum measurements on identical copies of quantum states $\rho$. For now, we restrict our discussions to projective measurements, so that in general, our measurement is described by a complete set of projectors $\{\Pi_m\}$, such that $\sum \Pi_m = I$. Often, one can associate the measurement to an observable $A$:

\begin{equation}\label{measurement operator}
    A = \sum_m m \Pi_m,
\end{equation}
where $\Pi_m$ is the projector onto the eigenspace of $A$ with eigenvalue $m$.
For each copy, we can make measurements using different observables, which corresponds to different sets of measurement operators.

\begin{figure}[h]
\centering
\includegraphics[scale = 0.5]{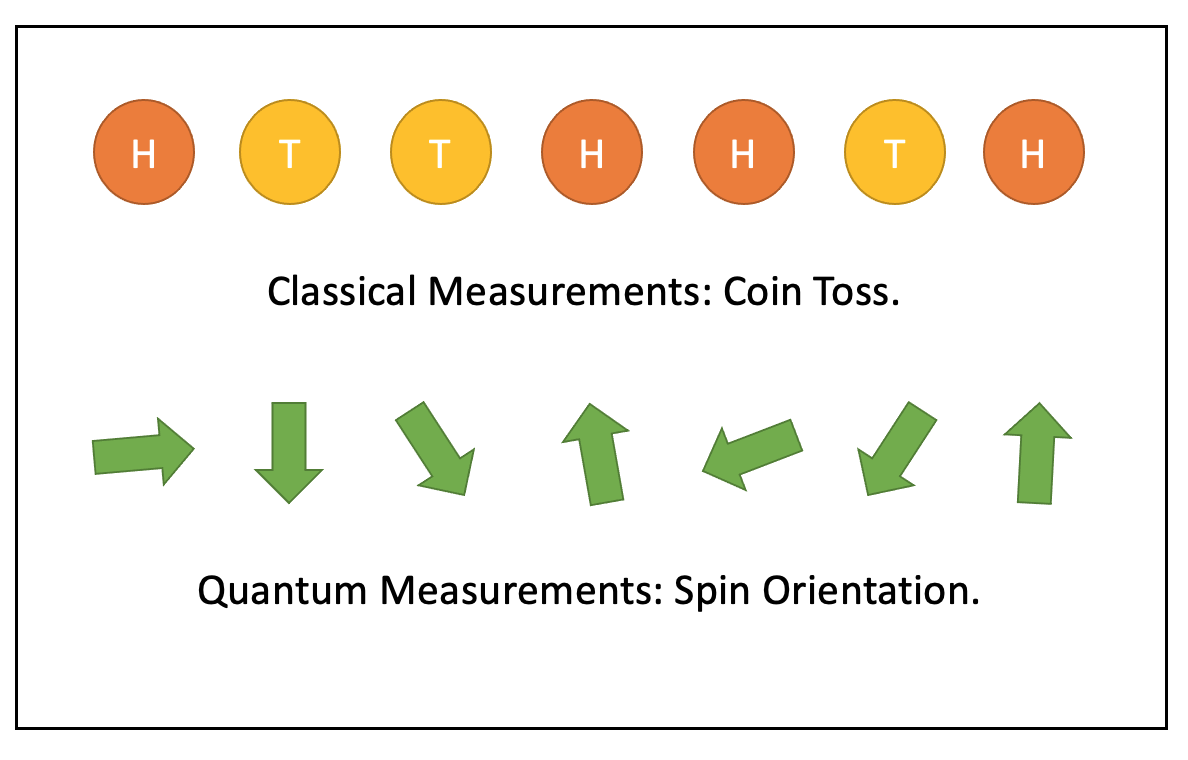}
\caption{Illustrations of classical and quantum measurements.}
\end{figure}

Constructing an empirical data distribution based on classical measurements is straightforward, and the corresponding cross entropy $H\big(P_D(x), P_{\theta}(x) \big) $ agrees with the average negative log-likelihood. Now we need to construct an empirical density matrix based on our quantum measurement results. However, the fact that the wave function collapses when making quantum measurements makes it impossible to make a simple and perfect reconstruction. We have to give up some information, which got lost when making the measurements. 
Hence, the question becomes: what information can we directly extract from a measurement, without solving equations or doing other complicated manipulations? 

We start from review the classical scenario and get inspirations from it.
When we flip a Bit and see it comes back with a "1", there are two ways of interpreting it, from a quantum perspective:

One way is to think that the "1" represents the eigenvalue $m$ of the observable $A$. Since what really matters is which eigenspace we project into, in general quantum cases, what we should care about is the corresponding projector $\Pi_m$. Hence, based on a single measurement outcome $m$, we record our data as $\Pi_m / \tr(\Pi_m)$. This perspective focuses more on the operator that is associated to the measurement as in Eq (\ref{measurement operator}), so we call it the operator perspective.

Another way is to think that the "1" represents the state after the measurement. The problem is that, we don't really know the true $\rho$, so we can only represent it by our model parameter $\sigma$, which is our belief of what $\rho$ should be. In this case, based on a single measurement outcome $m$, we record our data as $\Pi_m \sigma \Pi_m/\tr(\Pi_m \sigma)$, which is the state after the measurement, supposing the original state was $\sigma$. This perspective focuses more on the state, so we call it the state perspective.

Note that when all the projectors are of rank $1$, then the two perspectives agree, because

\begin{equation}\label{rank1equiv}
    \frac{\Pi_m \sigma \Pi_m}{\tr(\Pi_m \sigma)} = \Pi_m
= \frac{\Pi_m}{\tr(\Pi_m)}.
\end{equation}

A physical understanding is that suppose $\Pi_m = |\psi\rangle \langle \psi|$, then the quantum state we get after projection is $|\psi\rangle \langle \psi|$. Note that in the classical case, the outcome of the measurement is always in a certain state, so there is no distinction between the two perspectives described above. 

Now that we have $N$ identical copies, we can actually use $N$ (possibly) different observables, and all we need to record, are the projectors corresponding to the measurement results of the given observables. Hence, for the $n$-th copy of the quantum states, we construct the corresponding density matrices as 

\begin{equation}
    \rho_n^O = \frac{\Pi_n}{\tr(\Pi_n)}, \ \ 
    \rho_n^S = \frac{\Pi_n \sigma \Pi_n}{\tr(\Pi_n \sigma)},
\end{equation}
where the first equation corresponds to the operator perspective, and the second equation corresponds to the state perspective.

A direct generalization of the classical case gives the corresponding "empirical density matrices":

\begin{equation}
    \rho^O = \frac{1}{N}\sum_{n = 1}^N 
    \frac{\Pi_n}{\tr(\Pi_n)}, \ \ 
    \rho^S = \frac{1}{N}\sum_{n = 1}^N 
    \frac{\Pi_n \sigma \Pi_n}{\tr(\Pi_n \sigma)}.
\end{equation}

The quantum cross entropy can be written as follows, for the operator and the state perspectives, respectively:

\begin{equation}
    S(\rho^O, \sigma) = -\tr(\rho^O \log \sigma), \ \ S(\rho^S, \sigma) = -\tr(\rho^S \log \sigma).
\end{equation}
The extensivity of the quantum cross entropy (Proposition \ref{extensivity}) ensures that one can measure $N$ copies of the system in parallel and collect the measurement result. In terms of formula, if we write $\hat \rho^{O}= \rho_1^{O}\otimes...\otimes\rho_n^{O}$ and $\hat \sigma=\sigma\otimes...\otimes\sigma$, then $S(\hat \rho^{O},\hat \sigma)=N S(\rho^O, \sigma)$. Similarly, for the state perspective, we have $S(\hat \rho^{S},\hat \sigma)=N S(\rho^S, \sigma)$.

The average log-likelihood of the outcomes of the total $N$ measurements is 

\begin{equation}
    l(\sigma)
    = \frac{1}{N} \sum_{n=1}^N \log P(\Pi_n|\sigma)
    = \frac{1}{N} \sum_{n=1}^N \log \tr(\Pi_n \sigma).
\end{equation}

Now we show that both quantum cross entropies are lower bounded by the average negative log-likelihood.
The inequality can be seen as a reflection of the information loss when making quantum measurements.
Hence, when we design full quantum machine learning algorithms, it is crucial to reduce the number of quantum measurements. Moreover, we should take advantage of the deferred measurement principle as much as possible, so that we can reduce the disturbance from quantum measurements in the midst of training phase.

\begin{theorem}\label{theoremlikelyhoodinequality}
    If we measure $N$ identical copies of quantum state $\rho$ and define the quantum cross entropies as above, then 

\begin{equation}\label{likelyhoodbound}
    S(\rho^O, \sigma) \geq -l(\sigma), \ \ 
    S(\rho^S, \sigma) \geq -l(\sigma).
\end{equation}

\end{theorem}

%Comment: RHS classical negative cross entropy. The classical cross entropy only reveals the classical part of information, hence, smaller than the quantum case.

\begin{proof}
This is an application of Property \ref{crossentropybound} of the quantum cross entropy.

\begin{equation}
    \tr (\rho \log \sigma ) \leq \log \tr(\rho \sigma) .
\end{equation}

To prove the operator perspective inequality,  for each $n$, we take $\rho=\rho_n^{O}$, so

\begin{equation}\label{Oproof}
    \tr(\rho_n^O \log \sigma) =
    \frac{\tr(\Pi_n \log \sigma)}{\tr(\Pi_n)} 
    \leq 
    \log \Bigg( 
    \frac{\tr(\Pi_n \sigma)}{\tr(\Pi_n)} \Bigg)
    \leq \log \tr (\Pi_n \sigma).
\end{equation}

We then sum over $n$, which gives
\begin{equation}
    S(\rho^O, \sigma) = - \frac{1}{N}\sum_{n = 1}^N
    \tr(\rho_n^O \log \sigma) \geq -\frac{1}{N} \sum_{n = 1}^N \log \tr(\Pi_n \sigma) = -l(\sigma).
\end{equation}

To prove the state perspective inequality, for each $n$, we take $\rho=\rho_n^S$, so
\begin{equation}\label{Sproof1}
    \tr(\rho_n^S \log \sigma) =
    \frac{\tr(\Pi_n \sigma \Pi_n \log \sigma)}{\tr(\Pi_n \sigma \Pi_n)} 
    \leq 
    \log \Bigg( 
    \frac{\tr(\Pi_n \sigma \Pi_n \sigma)}{\tr(\Pi_n \sigma \Pi_n)} \Bigg).
\end{equation}

Let $A = \Pi_n \sigma \Pi_n$. The Schatten norm $||A||_1 \geq ||A||_2$. Since $A$ is a positive matrix,

\begin{equation}
    ||A||_1 = \tr |A| = \tr A \geq  \|A\|_2 = \big( 
    \tr|A|^2 \big)^\frac{1}{2} = \big( 
    \tr A^2 \big)^\frac{1}{2} .
\end{equation}

We then have
\begin{equation}\label{Sproof2}
    \log \Bigg( 
    \frac{\tr(\Pi_n \sigma \Pi_n \sigma)}{\tr(\Pi_n \sigma \Pi_n)} \Bigg)
    = \log \frac{\tr A^2}{\tr A} \leq \log \tr A 
    = \log \tr (\Pi_n \sigma).
\end{equation}

Summing over $n$ gives 

\begin{equation}
    S(\rho^S, \sigma) = - \frac{1}{N}\sum_{n = 1}^N
    \tr(\rho_n^S \log \sigma) \geq -\frac{1}{N} \sum_{n = 1}^N \log \tr(\Pi_n \sigma) = -l(\sigma).
\end{equation}

\end{proof}

\begin{proposition}[Equality condition of Theorem~\ref{theoremlikelyhoodinequality}]\label{theoremequalityremark} Let $P_{\sigma}$ 
 be the support projector of state $\sigma$, and $\{\Pi_n\}$ be a set of measurement projectors corresponding to an observable $\hat O$. Then
$S(\rho^O, \sigma)=-l(\sigma)$ if and only if for all $\Pi_n$'s such that $\Pi_n  P_{\sigma}\neq 0$, $\text{rank}~(\Pi_n) = 1$ and $[\Pi_n,\sigma]=0$. $S(\rho^S, \sigma)=-l(\sigma)$ if and only if for all~$\Pi_n$'s such that~$\Pi_n  P_{\sigma}\neq 0$, $\text{rank}~(\Pi_n P_{\sigma}) = 1$ and $[\Pi_n,\sigma]=0$.
\end{proposition}

\begin{proof}
We only consider $\Pi_n$'s with $\Pi_n  P_{\sigma}\neq 0$, because otherwise they will not contribute to the sum in $S(\rho^O,\sigma)$ or $S(\rho^S,\sigma)$.
For all $\Pi_n$'s such that $\Pi_n  P_{\sigma}\neq 0$, the last inequality in Eq.\eqref{Oproof} achieves equality if and only if $\text{rank}(\Pi_n) = 1$. The first inequality in Eq.\eqref{Oproof} achieves equality if and only if $[\Pi_n,\sigma]=0$. They combined are the equality conditions for $S(\rho^O, \sigma)=-l(\sigma)$.

The Schatten norm inequality Eq.\eqref{Sproof2}  achieves equality if and only if $\text{rank}(\Pi_n \sigma \Pi_n) = 1$ for all $\Pi_n \sigma \Pi_n\neq 0$. This is equivalent to $\text{rank}(\Pi_n P_{\sigma}) = 1$ whenever $\Pi_n P_{\sigma}\neq 0$. This is proved as Lemma \ref{rankequal}. The inequality Eq.\eqref{Sproof1} requires $[\Pi_n \sigma \Pi_n,\sigma]=0$. In the case $\text{rank}(\Pi_n P_{\sigma})=1$, $[\Pi_n \sigma \Pi_n,\sigma]=0$ is equivalent to $[\Pi_n ,\sigma]=0$. This equivalence is proved as Lemma \ref{commutatorequal}. Both lemmas are collected in Appendix \ref{append1}.
\end{proof}

The equality condition has direct physical meanings. If we measure an observable $\hat O$, then $S(\rho^O, \sigma)=-l(\sigma)$ holds if and only if $\sigma$ has support only on the non-degenerate part of the eigenspace of $\hat O$ and $[\hat O,\sigma]=0$. This means that if we measure $\hat O$ in the state $\sigma$, the outcome is classical, in the sense that eigenstates are non-degenerate and have no correlations among each other. The equality condition for $S(\rho^S, \sigma)=-l(\sigma)$ is almost the same yet slightly weaker, in the sense that it allows the state $\sigma$ to take support on degenerate eigenspaces of the operator $\hat O$, but only 1 dimensional. Intuitively, if we measure the operator $\hat O$ in the state $\sigma$, it again looks classical. For the support on the non-degenerate eigenspaces of $\hat O$, they are the same case as in the state perspective. For the support on the degenerate eigenspaces, since it only allows 1 dimensional subspace in each degenerate eigenspace, we cannot tell whether it is degenerate or not by looking at the outcome of the measurement in this state $\sigma$, so it also looks classical.

We discussed around Eq (\ref{rank1equiv}) that the two perspectives give identical results when the projector is of rank 1, which corresponds to the non-degenerate eigenspaces of the operator $\hat O$ we measure. The distinction lies in the different treatments of the degenerate eigenspaces. In the operator perspective, we interpret the post-measurement state as a maximally mixed state supported on this subspace. While in the state perspective, we just project to this subspace, retaining the original information of the state $\sigma$ in this subspace.

\subsection{Violation of Theorem \ref{theoremlikelyhoodinequality} for POVM}

A natural question is if we can generalize the discussion of this section from projective measurements to POVM's and establish a generalized version of Theorem \ref{theoremlikelyhoodinequality}. Straightforward generalization is not always correct. For example, in the state perspective, consider a qubit state $\sigma=\begin{pmatrix}2/3&0\\0&1/3\end{pmatrix}$ and POVM operators $M_1=|0\rangle\langle0|=\begin{pmatrix}1&0\\0&0\end{pmatrix}$, $M_1=|0\rangle\langle1|=\begin{pmatrix}0&1\\0&0\end{pmatrix}$. Indeed $M_1^\dagger M_1 +M_2^\dagger M_2=I_2$. Now we can compute explicitly that $\tr(\rho_1^s\log\sigma)=\log\tr(M_1^\dagger M_1\sigma)=\log 2/3$, $\tr(\rho_2^s\log\sigma)=\log 2/3$, while $\log\tr(M_2^\dagger M_2\sigma)=\log 1/3$. Here $\rho_n^s$ is the post-measurement state $\rho_n^s=M_n \sigma M_n^\dagger /\tr(M_n \sigma M_n^\dagger)$. If we measure $N$ times and get the quantum cross entropy and the average negative log-likelihood, the analogy of the inequality Eq (\ref{likelyhoodbound}) does not hold. 

The mathematical origin of this counterexample is that the POVM operator $M_2$ is different from the usual projective measurement operator $P_2=|1\rangle\langle1|$ via a unitary, $M_2=U P_2$. This does not affect the probability of the outcome $\tr(M^\dagger_2 M_2 \sigma)$ but will indeed change the post-measurement state comparing to the projective measurement case $\rho_2^s=M_2 \sigma M_2^\dagger /\tr(M_2 \sigma M_2^\dagger)=U P_2 \sigma P_2 U^\dagger /\tr(P_2 \sigma)$. When calculating the quantum cross entropy, this unitary effectively acts on $\sigma$ hence changes the reference state.

In this counterexample, the post-measurement states of the system $\rho_1^s$ and $\rho_2^s$ are identical. The distinction between the measurement result 1 and 2 lies in the post-measurement state of the ancilla not the system. This suggests that when studying POVM measurements, only focusing on the post-measurement state of the system is not sufficient.

\section{Conclusion}

In this paper, we defined the quantum cross entropy. We showed its positivity, invariance under unitaries, linearity on the first argument, convexity on the second argument, and joint convexity. Moreover, we proved its lower bounds, which is related to the quantum fidelity. We investigated its relation to maximum likelihood principle, a driven force of the developments of many machine learning algorithms. Since our ultimate goal is to achieve full quantum machine learning, we explored different notions of quantum data, upon which we constructed the quantum cross entropy.
We first showed that minimizing the quantum cross entropy undisturbed by quantum measurements is equivalent to maximizing likelihood. We then examined the case when the quantum cross entropy is reconstructed from the empirical density matrix based on measurement outcomes, where the quantum cross entropy is lower-bounded by negative log-likelihood. The differences between two scenarios mainly come from whether the quantum cross entropy is disturbed by quantum measurements or not. We conclude that to enhance the performances of full quantum machine learning, it is crucial to apply the deferred measurement principle, so that we can use the undisturbed, better-behaved quantum cross entropy. For future work, we aim at designing quantum algorithms which utilize the quantum cross entropy, and demonstrating its quantum supremacy.

\section*{Acknowledgement}
Z.S. was inspired by Stanford's machine learning course. Z.S. would like to thank Yuri Lensky for an inspirational discussion on measurement operators. We would like to thank Adam Brown, Tarun Grover, Zhaoyu Han, Patrick Hayden, Richard Kueng, and Brian Swingle for helpful comments. In particular, we appreciate detailed comments on an earlier draft from Brian Swingle. Z.S. is supported by the Simons Foundation. Y.W. is supported in part by the U.S. Department of Energy, Office of Science, Office of Advanced Scientific Computing Research, Accelerated Research in Quantum Computing (FAR-QC) and by the Air Force Office of Scientific Research under award number FA9550-19-1-0360.

\begin{appendix}
\section{Supplementary proofs of Proposition \ref{theoremequalityremark}} \label{append1}

\begin{lemma}\label{rankequal}
Let $\Pi_n$ be a projector and $\sigma$ a density operator with $P_{\sigma}$ its support projector, then $\text{rank}~(\Pi_n P_{\sigma})=\text{rank}~(\Pi_n\sigma\Pi_n)$.
\end{lemma}
\begin{proof}
On the one hand, $\Pi_n\sigma\Pi_n=\Pi_n P_{\sigma}\sigma P_{\sigma} \Pi_n$, this implies $\text{rank}(\Pi_n\sigma\Pi_n)\leq\text{rank}(\Pi_n P_{\sigma})$. On the other hand, we can make use of the inequality
\begin{equation*}
\text{rank}(AB)+\text{rank}(BC)\leq\text{rank}(B)+\text{rank}(ABC).
\end{equation*}
Let $A=\Pi_n$, $B=P_{\sigma}$ and $C=\sigma^{1/2}$. Note that $P_{\sigma}\sigma^{1/2}=\sigma^{1/2}$ and $\text{rank}(P_\sigma)=\text{rank}(\sigma^{1/2})$, the above inequality implies
\begin{equation*}
    \text{rank}(\Pi_n P_{\sigma})\leq \text{rank}(\Pi_n\sigma^{1/2})=\text{rank}((\Pi_n\sigma^{1/2})(\Pi_n\sigma^{1/2})^{\dagger})=\text{rank}(\Pi_n\sigma\Pi_n).
\end{equation*}
Hence, $\text{rank}(\Pi_n P_{\sigma})=\text{rank}(\Pi_n\sigma\Pi_n)$.
\end{proof}

\begin{lemma}\label{commutatorequal}
Let $\sigma$ be a density operator and $\Pi_n$ a projector. If $\text{rank}~(\Pi_n\sigma\Pi_n)=1$, then $[\Pi_n \sigma \Pi_n,\sigma]=0$ is equivalent to $[\Pi_n ,\sigma]=0$.
\end{lemma}
\begin{proof}
From Lemma \ref{rankequal}, $\text{rank}(\Pi_n\sigma)=1$. So the singular value decomposition can be written as $\Pi_n\sigma=U S V^{\dagger}$, where $S=\text{diag}(s,0,0,...)$ is a rank 1 diagonal matrix with only first element non-zero, and $U, V$ are unitaries. The condition $[\Pi_n \sigma \Pi_n,\sigma]=0$ can be written as 
\begin{equation*}
    U S V^{\dagger} U S V^{\dagger}= V S U^{\dagger} V S U^{\dagger}.
\end{equation*}
Defining $W=V^\dagger U$, the above equation can be rewritten as
\begin{equation*}
    W S W S W= SW^\dagger S.
\end{equation*}
This sets constraints to the form of the unitary $W=\begin{pmatrix}w&\vec 0\\ \vec 0 { }^T& W^\prime\end{pmatrix}$, with $w^3=w^*$ and $W^\prime$ an arbitrary unitary of one lower dimension. So $w$ can take $\pm 1, \pm i$. Plugging $U=V W$ back into $\Pi_n \sigma=U S V^\dagger$, we find $\Pi_n \sigma=(w/w^*)\sigma \Pi_n$. $w=\pm1$ leads to $[\Pi_n,\sigma]=0$ and $w=\pm i$ leads to $\{\Pi_n,\sigma\}=0$. However, if $\{\Pi_n,\sigma\}=0$, then on the one hand $\sigma \Pi_n=-\Pi_n \sigma$, on the other hand $\sigma \Pi_n \Pi_n=-\Pi_n \sigma \Pi_n=\Pi_n \Pi_n \sigma=\Pi_n \sigma$. This implies $\Pi_n\sigma=0$, which contradicts with $\text{rank}(\Pi_n\sigma)=1$. So the only possibility is $[\Pi_n,\sigma]=0$.

The other direction is manifest.
\end{proof}

\end{appendix}

\bibliographystyle{unsrt}
\bibliography{ref}

\end{document}